\documentclass[prb,twocolumn,showpacs]{revtex4}
\usepackage{graphicx}
\usepackage{amsfonts}
\usepackage{amsmath}
\usepackage{amssymb}
\usepackage{bm}
\usepackage{hyperref}
\usepackage{slashed}

\begin{document}

\title{D-wave superconductivity induced by proximity to the non-uniform chiral spin liquid on a square lattice}

\author{Rui Wang$^{1,2}$}
\author{Haiyuan Zou$^{4}$}

\author{Tigran Sedrakyan$^{5}$}

\author{Baigeng Wang$^{1,3}$}
\email{bgwang@nju.edu.cn}

\author{D. Y. Xing$^{1,3}$}

\affiliation{$^1$Department of Physics, Nanjing University, Nanjing 210093, China
}
\affiliation{$^2$Department of Physics and Astronomy, Shanghai Jiao Tong University, Shanghai 200240, China
}
\affiliation{$^3$National Laboratory of Solid State Microstructures and Collaborative Innovation Center of Advanced Microstructures, Nanjing University, Nanjing 210093, China
}

\affiliation{$^4$Tsung-Dao Lee Institute, Shanghai Jiao Tong University, Shanghai 200240, China
}
\affiliation{$^5$Department of Physics, University of Massachusetts Amherst, Amherst, Massachusetts 01003, USA}

\date{\today }

\begin{abstract}
We use the tensor network algorithm to show evidences of a non-uniform  chiral spin-liquid (CSL) ground state in a frustrated spin-$1/2$ model on a square lattice, in the regime of moat-like band structure of the lattice, i.e., a band with infinitely degenerate energy minima attained along a closed contour in the Brillouin zone.
The analytical description of the state is given by the effective field theory of a topological square-lattice fermionic Chern insulator coupled to the Chern-Simons gauge field.
The observed non-uniform CSL has a substantial effect on a nearby free-fermion environment.
We show that, in the presence of arbitrarily small spin exchange interaction, the CSL can endow a gauge-field-modulated effective interaction between the environmental fermions. The induced effective interaction can be attractive within a significant parameter region, leading to an instability towards d-wave superconductivity in the fermionic bath.
\end{abstract}

\pacs{}
\maketitle

\emph{Introduction.}--
Superconductivity, discovered by Kamerlingh and Onnes in 1911, is one of the paradigmatic phases of strongly interacting quantum many-electron systems.
Variety of mechanisms of the pairing of electrons were extensively studied over decades, since the first experimental observation. Much of our understanding of the phenomenon comes from the celebrated Bardeen-Cooper-Schrieffer (BCS) theory developed in 1957 \cite{Bardeen}.
For traditional BCS superconductors (SCs), it is well accepted that the electron-phonon interaction is responsible for the formation of Cooper pairs. For high-$T_c$ superconductors, though still being debated, many different theories of the pairing mechanism have been put forward \cite{palee,Gros,Paramekanti,Sorella,Shih,White,Jarrel,Senechal,Tremblay,Eder,Anderson,zZou,Andersona,Kivelsonn,Andersonc,srwhite,fczhang,haldane,Affleckk,Baskarann,Andersondd,Chakravarty,Emery,Miyake,Hirsch,Monthonx,Kyung,Dahm,Monthouxx,Scalapino,Abanov,vjemery,gchen,Mishchenko,Barone}.

Another fascinating example of complex phases of strongly interacting quantum many-body systems is the quantum spin liquid -- a Mott-insulator with strong quantum fluctuations that suppress ordering down to the lowest temperatures. A characteristic property of quantum spin-liquids is that
they support  fractionalization of low-lying excitations while their analytical description implies the formation of gauge fields and a topological order
of the low-energy effective theories.

An interesting though exotic theoretical framework for the interplay of quantum spin-liquids and superconductivity was proposed in Refs.~\onlinecite{Anderson,Andersona,Fazekas}
in the context of high-$T_c$ systems, where the origin of the $U(1)$ gauge field was related to the spin-overlap on adjacent $\mathrm{Cu}$ atoms in the $\mathrm{CuO}_2$ planes. This is the so-called resonating valence-bond picture of cuprates, where superconductivity was proposed to be triggered upon carrier doping of the disordered magnet\cite{Konika,rbLaughlin,Rokhsar,Kotliar,kyyang,Gabriel,Chatterjee}.

In this paper, we consider the interplay between a quantum spin-liquid with spontaneously broken time-reversal symmetry (TRS) and superconductivity, yet from a distant perspective.
A key representative of such spin-liquids is the CSL \cite{xgwenb,ssgong,Thomale,Schulz,Sindzingre,jfyu,lwangb,akiaev,Mkhitaryan,hyao,Sedrakyann,Messio,Bauer,kkumar,Wietek,Nataf,Bierii,wjh,Panfilov,
Hickey,Hickeyy,Bieri,Wietekk,tsvelik,yzhou,Savary,Halimeh,Kalmeyer,ssgongb,yche,ycheb,Varney,Carrasquilla,Ciolo,zzhu}, which is close in its nature to the quantum Hall state: it is gapped in bulk, and simultaneously supports massless chiral edge excitations.
Deviating from the standard approach to similar problems, where one considers the effect of carriers on the spin system\cite{amtsvelik}, we ask the opposite question: if one starts with a CSL state and couples it to an external, free-electron bath, how will the state of the Mott-insulator affect the free-electron environment?

The focus of the present article is twofold. Firstly, we numerically show that a non-uniform CSL\cite{sedrakyanb} -- a state where Ising antiferromagnetic (AFM) order coexists with CSL\cite{Varney,Carrasquilla,Ciolo,zzhu,sedrakyana,sedrakyanb,sedrakyanc} -- can be stabilized on a square lattice as the ground state of a spin-1/2 XY model with competing interactions.  This happens in the broad parameter regime where the lattice band structure exhibits an energy minimum along a closed line -- dubbed {\em the moat} -- in the Brillouin zone.  As the next step, we identify the low-energy effective field theory describing the state, that can be understood as a square-lattice fermionic Chern insulator coupled to the Chern-Simons gauge field.
It is shown that the mean-field treatment of the topological field theory yields results which are in a good agreement with the numerically observed CSL, in terms of the phase diagram, the Ising order, and the chirality order $\chi_{\mathbf{r},\mathbf{r}^{\prime},\mathbf{r}^{\prime\prime}}=\langle\mathbf{S}_{\mathbf{r}}\cdot(\mathbf{S}_{\mathbf{r}^{\prime}}\times\mathbf{S}_{\mathbf{r}^{\prime\prime}})\rangle$ \cite{xgwenb}.
Secondly, we demonstrate the existence of an unexpected ``proximity effect" of the non-uniform CSL, when it is coupled with a free-electron bath.
Even at arbitrarily small spin exchange coupling, the CSL can induce pairing of electrons in the environmental bath, giving rise to a $d$-wave superconducting state for a significant parameter region. The pairing of electrons is attributed to the CS gauge field inherited from the CSL. Similar to the classical picture where two opposite magnetic fluxes generate opposite in-plane cyclotron motion of two electrons and thus an attractive Ampere force between them, the staggered CS gauge field from CSL can result in an effectively attractive interaction between electrons and thus a superconducting state in the bath. Furthermore, gapless nodes of the SC gap with sign changes are found along the Fermi surface, indicating a $d_{x^2-y^2}$ pairing symmetry. We show that this is mainly due to the square lattice symmetry as well as the $\pi$-flux gauge field inherited by the non-uniform CSL from the planar N\'{e}el AFM parent state at the weak frustration regime.

\begin{figure}
\includegraphics[width=3.4in]{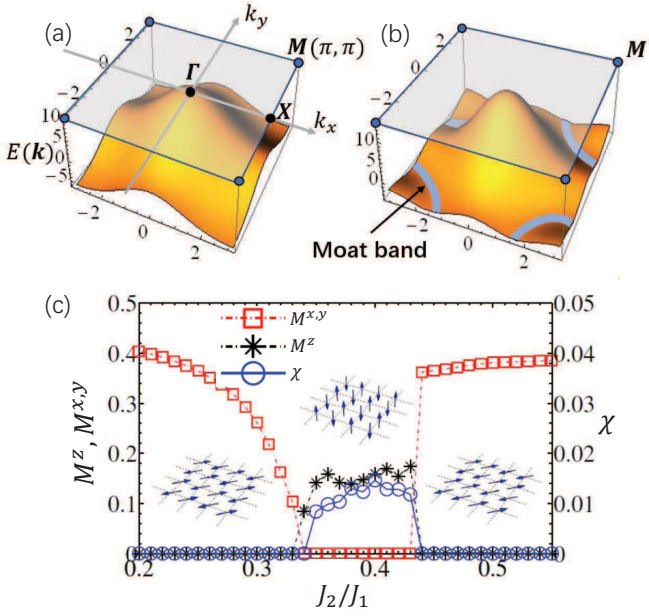}
\label{fig5b}
\caption{(color online). The typical hardcore boson spectrum for (a) $J_2/J_1<0.25$  and (b) $J_2/J_1>0.25$. (c) Tensor network results of $M^{x,y}\equiv \sum_{\mathbf{r}} \sqrt{\langle S^{x}_{\mathbf{r}}\rangle^2+\langle S^{y}_{\mathbf{r}}\rangle^2}/N$, $M^{z}\equiv \sum_{\mathbf{r}} |\langle S^{z}_{\mathbf{r}}\rangle|/N$, and the chirality order $\chi$ as a function of $J_2/J_1$ at virtual dimension $D=4$. Insets schematically show the spin orders for the three phases. $\chi$ is plot with a weak perturbation  \cite{explain} $\beta=0.01$.}
\end{figure}

\emph{Model Hamiltonian}.-- The total Hamiltonian under consideration consists of the frustrated spin $s=1/2$ system,  the fermionic environment, and a coupling term between them,
$H_{tot}=H_{sys}+H_{env}+H_c$. For the spin system
we consider the $J_1$-$J_2$-$J_3$ AFM XY model on the square lattice:
\begin{equation}\label{eq1}
\begin{split}
  H_{sys}&=\sum_{\mathbf{r},\mathbf{r}^{\prime}}J_{\mathbf{r},\mathbf{r}^{\prime}}(S^x_{\mathbf{r}}S^x_{\mathbf{r}^{\prime}}+S^y_{\mathbf{r}}S^y_{\mathbf{r}^{\prime}}),
\end{split}
\end{equation}
with the nearest-neighbor (NN) interaction $J_{\langle\mathbf{r},\mathbf{r}^{\prime}\rangle}=J_1$, next-NN (NNN) $J_{\langle\langle\mathbf{r},\mathbf{r}^{\prime}\rangle\rangle}=J_2$, and next-NNN (NNNN) $J_{\langle\langle\langle\mathbf{r},\mathbf{r}^{\prime}\rangle\rangle\rangle}=J_3$. The Hamiltonian
$H_{env}$ describes a free environmental bath with spinful fermions, $\{c^{(\dagger)}_{\mathbf{r}\sigma}\}$, $\sigma=\uparrow,\downarrow$, hopping on the square lattice:
 \begin{equation}\label{eqadd2}
H_{env}=-t\sum_{\langle\mathbf{r},\mathbf{r}^{\prime}\rangle}c^{\dagger}_{\mathbf{r}\sigma}c_{\mathbf{r}^{\prime}\sigma}+h.c.
 \end{equation}
The latter is assumed to be coupled  to the spin system $H_{sys}$ through the XY spin-exchange interaction
\begin{equation}\label{eq2}
  H_{c}=J_c\sum_{\mathbf{r}}c^{\dagger}_{\mathbf{r}\alpha}(S^{x}_{\mathbf{r}}\sigma^x_{\alpha\beta}+S^{y}_{\mathbf{r}}\sigma^y_{\alpha\beta})c_{\mathbf{r}\beta},
\end{equation}
where $\boldsymbol{\sigma}$ is the Pauli matrix denoting the spin degrees of freedom of c-fermions. Throughout this work, we assume the weak coupling condition with $J_c\ll t, J_i$, ($i=1,2,3$). The model $H_{tot}$ offers a unique opportunity to both explore the frustration-induced stabilization of the CSL and an unambiguous signal of formation of d-wave superconductivity in the free-fermion bath.

\emph{Non-uniform CSL on a square lattice}.-- We firstly study $H_{sys}$, with $J_c$ being "turned off". Using the hardcore boson representation of spin-1/2 operators \cite{explain}, we obtain the single-particle spectrum, as shown in Fig.1(a)-(b), where $J_3=J_2/2$. It is found that for $J_2/J_1<0.25$, the energy minimum of the spectrum resides at the $M$-point in the reciprocal space, while once $J_2/J_1 > 0.25$, an infinitely degenerate nodal loop emerges, consisting of energy minima surrounding the $M$-point\cite{footnote-here}.

Near the moat bottom energy, $E_c$, the single-particle density of states diverges as $(E-E_c)^{-1/2}$, resembling  the one-dimensional (1D) Tonks-Girardeau case\cite{tonks,Girardeau,Lieb,cnyang}, motivating the possibility of statistical transmutation and absence of  in-plane magnetic orders \cite{sedrakyana}.
This suggests a possible nontrivial ground state for $J_2> 0.25$. To probe the many-body nature of the ground state, we use the tensor-network ansatz~\cite{Verstraete,Cirac,Vidal}, in which, the many-body wavefunction is constructed by local tensors, and the quantum entanglement is controlled by the virtual bond-dimension $D$ of these tensors. We apply an update \cite{Jiang,TEBD} based on the Trotter-Suzuki expansion, and a real space coarse-graining procedure~\cite{Xie}, known as the higher-order tensor renormalization group~\cite{footnote}, for the evolution of ground state  and the calculation of order parameters.

The ground state expectation of spin operators, $\langle S^i_{\mathbf{r}}\rangle$, with $i=x,y,z$ are obtained, which suggest three different phases with varying $J_2/J_1$. As shown in Fig.1(c), for $0.33\gtrsim J_2/J_1$, the expected planar N\'{e}el AFM order is found with $\langle S^x_{\mathbf{r}}\rangle\neq 0$ and $\langle S^z_{\mathbf{r}}\rangle=0$, $\forall\mathbf{r}$. For $0.44\gtrsim J_2/J_1\gtrsim 0.33$, the planar spin ordering vanishes,
stabilizing a phase that is completely disordered in the XY-plane with $\langle S^{x,y}_{\mathbf{r}}\rangle=0$, $\forall\mathbf{r}$. Moreover, we observe an {\em unexpected} out-of-plane N\'{e}el AFM Ising ordering with $\langle S^z_{\mathbf{r}}\rangle\neq0$, $\forall\mathbf{r}$. For $J_2/J_1\gtrsim 0.44$,
the out-of-plane Ising order is again replaced by a planar magnetic ordering with the nesting vector $(\pi/2,\pi)$.
What is even more intriguing, with introducing a weak TRS-breaking phase-perturbation to $H_{sys}$, namely by replacing the NN interaction term by $H_1\rightarrow J_1\sum_{\langle\mathbf{r},\mathbf{r}^{\prime}\rangle}\left(e^{\pm i\beta\pi}S^{+}_{\mathbf{r}}S^{-}_{\mathbf{r}^{\prime}}+h.c.\right)$ \cite{Plekhanov} with an infinitesimal $\beta$, we obtain a strong chirality order $\chi_{\mathbf{r},\mathbf{r}^{\prime},\mathbf{r}^{\prime\prime}}$ in the region $0.44\gtrsim J_2/J_1\gtrsim 0.33$, with $\mathbf{r},\mathbf{r}^{\prime},\mathbf{r}^{\prime\prime}$ being the three NN sites in the elemental plaquette. In sharp contrast,  $\chi_{\mathbf{r},\mathbf{r}^{\prime},\mathbf{r}^{\prime\prime}}$ always remains zero in the ordered states at $0.33\gtrsim J_2/J_1$ and $J_2/J_1\gtrsim 0.44$ even in the presence of a finite $\beta$.  Furthermore, a linear scaling behavior of $\chi_{\mathbf{r},\mathbf{r}^{\prime},\mathbf{r}^{\prime\prime}}$ with $\beta$ is found in $0.44\gtrsim J_2/J_1\gtrsim 0.33$, which is an inherent property of CSL \cite{explain,explain2}. The spontaneous TRS breaking is further supported by our density matrix renormalization group calculations on a cylinder geometry, which identifies the  interchange of ground state between two degenerate topological sectors with the adiabatic evolution of the twisted angle \cite{yche} from $0$ to $2\pi$ \cite{footnote}.
The above observations suggest the emergence of a chiral state which (a) is disordered in XY plane (b) has an out-of-plane Ising ordering  (c) spontaneously breaks TRS \cite{sedrakyana,sedrakyanb,sedrakyanc}.

To describe these findings analytically, we fermionize the hard-core bosons using the Chern-Simons flux attachment \cite{sedrakyana,sedrakyanc}, similar to the approaches in fractional quantum Hall (FQH) systems. The lattice version of CS fermionization \cite{jain,alopez,Halperin,rshankar,jackson} is formulated as $S^{\pm}_{\mathbf{r}}=f^{\pm}_{\mathbf{r}}\hat{U}^{\pm}_{\mathbf{r}}$, with the string operator $\hat{U}^{\pm}_{\mathbf{r}}=e^{\pm i\sum_{\mathbf{r}^{\prime}\neq\mathbf{r}}\mathrm{arg}[\mathbf{r}-\mathbf{r}^{\prime}]n_{\mathbf{r}^{\prime}}}$ and the spinless CS fermion operator $f_{\mathbf{r}}$. The nonlocal string operators ensure the $\mathrm{SU}(2)$ algebra of the spins. In this representation, nearby string operators will combine to generate the CS gauge field  $\mathbf{A}_{\mathbf{r}}=\sum_{\tilde{\mathbf{r}}}n_{\tilde{\mathbf{r}}}\nabla_{\mathbf{r}}\mathrm{arg}(\mathbf{r}-\tilde{\mathbf{r}})$.
Similarly, the CS phases are defined as $A_{\mathbf{r},\mathbf{r}^{\prime}}=\sum_{\tilde{\mathbf{r}}} n_{\tilde{\mathbf{r}}}[\mathrm{arg}(\mathbf{r}-\tilde{\mathbf{r}})-\mathrm{arg}(\mathbf{r}^{\prime}-\tilde{\mathbf{r}})] $. In the following, we introduce the notation $A^{l,l^{\prime}}_{\mathbf{r},\mathbf{r}^{\prime}}$ (with $l,\l^{\prime}=a,b$) to denote the CS phases on bond $\langle\mathbf{r},\mathbf{r}^{\prime}\rangle$ and from sublattice $l$ to $l^{\prime}$.  Let us note in passing that the number operator of CS fermions satisfies $n_{\mathbf{r}}=S^z_{\mathbf{r}}+1/2$.

The generated CS gauge field $\mathbf{A}_{\mathbf{r}}$ obeys the ``Stocks theorem" \cite{sedrakyanb}, $B_{\mathbf{r}}=\oint d\mathbf{r}\cdot\mathbf{A}_{\mathbf{r}}=2\pi n_{\mathbf{r}}$. To implement this constraint condition \cite{rui}, we introduce a Lagrangian multiplier $A^0_{\mathbf{r}}$. $H_{sys}$ is then mapped to a field theory describing spinless fermions coupled to the CS gauge field,
\begin{equation}\label{eqn4}
\begin{split}
  S_{sys}&=\int dt[\sum_{\mathbf{r}}f^{\dagger}_{\mathbf{r}}(i\partial_t-A^0_{\mathbf{r}})f_{\mathbf{r}}+\frac{1}{2\pi}\sum_{\mathbf{r}}A^0_{\mathbf{r}}B_{\mathbf{r}}\\
  &-\sum_{\mathbf{r},\mathbf{r}^{\prime}}J_{\mathbf{r},\mathbf{r}^{\prime}}f^{\dagger}_{\mathbf{r}}f_{\mathbf{r}^{\prime}}e^{iA_{\mathbf{r},\mathbf{r}^{\prime}}}-h.c.].
\end{split}
\end{equation}
The bipartite spin model allows for two different sublattices a, b. The sublattice degrees of freedom, in the fermionic representation, are translated as the staggered NN CS gauge fields \cite{rui}, indicated by the yellow arrows in Fig.2(b), where we work under the gauge that the NN phases only reside on horizontal bonds, with $A^{a,b}_{\mathbf{r},\mathbf{r}+\mathbf{e}_x}=A^{a,b}_{\mathbf{r},\mathbf{r}-\mathbf{e}_x}=-A^{b,a}_{\mathbf{r},\mathbf{r}+\mathbf{e}_x}=-A^{b,a}_{\mathbf{r},\mathbf{r}-\mathbf{e}_x}=\phi/2$, where $\mathbf{e}_x$, $\mathbf{e}_y$ are the lattice vectors of the square lattice.
Guided by the numerical observation of the Ising ordering in $0.44\gtrsim J_2/J_1\gtrsim 0.33$ which is translated as the density wave order of the CS fermions with $n^A_{\mathbf{r}}-n^B_{\mathbf{r}}\neq0$, we shall look for mean-field states allowing for broken symmetry between A and B sublattices. We restrict ourselves to the simplest but general choice with $A^{a,a}_{\mathbf{r},\mathbf{r}+\mathbf{e}_x+\mathbf{e}_y}=A^{b,b}_{\mathbf{r},\mathbf{r}+\mathbf{e}_x-\mathbf{e}_y}=\gamma_2$, $A^{a,a}_{\mathbf{r},\mathbf{r}+\mathbf{e}_x-\mathbf{e}_y}=A^{b,b}_{\mathbf{r},\mathbf{r}+\mathbf{e}_x+\mathbf{e}_y}=\gamma_1$, as shown by the orange and green arrows in Fig.2(b) respectively. These NNN CS phases generate the staggered CS fluxes in a unit cell in Fig.2(a), with $\Delta\gamma=\gamma_2-\gamma_1$ and $-\Delta\gamma$ in the left and right triangle, respectively.


\begin{figure}[tbp]
\includegraphics[width=\linewidth]{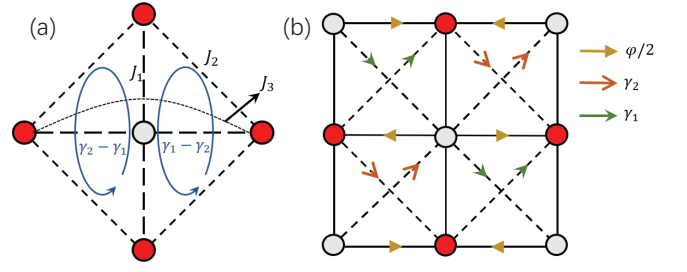}
\caption{(color online). (a) The unit cell with a (white site) and b (red site) sublattice, and with NN ($J_1$), NNN ($J_2$), NNNN $(J_3)$ interaction. CS fluxes associated with spontaneous TRS breaking are shown. (b) The NN, NNN CS gauge fields on the square lattice.}
\end{figure}

\begin{figure}[b]
\includegraphics[width=\linewidth]{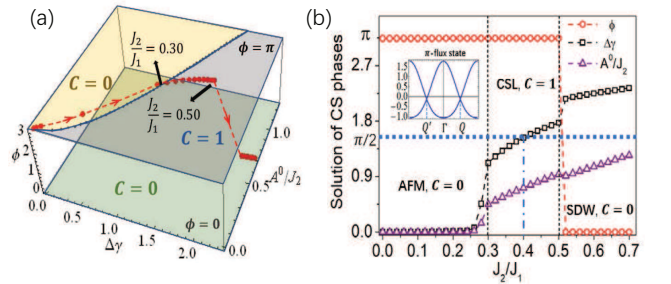}
\label{fig5a}
\caption{(color online) (a) The self-consistently calculated trajectory of CS phases with increasing $J_2/J_1$. (b) $\phi$, $\Delta\gamma$ and $A^0$ versus $J_2/J_1$ . The inset shows the CS fermion spectrums for the $\pi$-flux state, which ends up with planar N\'{e}el AFM order with further taking into account the fluctuation of gauge field. }
\end{figure}

The CS fermions are Gaussian in Eq.\eqref{eqn4} and one can integrate them out to obtain the fermionic free energy functional, $S_{eff}=W[A^0,\phi,\gamma_1,\gamma_2]+\frac{1}{2\pi}\sum_{\mathbf{r}}A^0B_{\mathbf{r}}[\phi,\gamma_1,\gamma_2]$,
where $A^0$ is the antisymmetric Lagrangian multiplier defined as combinations of $A^0_{\mathbf{r}}$ at a, b sites in a unit cell $A^0=\left(A^0_{\mathbf{r},a}-A^0_{\mathbf{r},b}\right)/2$, and the symmetric component has been gauged out at half-filling \cite{sedrakyanb}. The free energy $W[A^0,\phi,\gamma_1,\gamma_2]=\sum_{\mathbf{k},n\in occ}E_n(\mathbf{k},A^0, \phi,\gamma_1,\gamma_2)$ with $E_n(\mathbf{k})$ being the energy of the n-band CS fermions and the sum over all occupied states.  By minimizing $S_{eff}$ with respective to the CS phases and $A^0$, we obtain the mean-field solution \cite{explain}  $(A^0,\Delta\gamma,\phi)$. The evolution of $(A^0,\Delta\gamma,\phi)$ upon increasing  $J_2/J_1$ is represented by the red dashed curve in Fig.3(a). $(A^0,\Delta\gamma,\phi)$ starts from the $(0,0,\pi)$ point in the 3D order parameter space, and moves along the direction of the arrows with gradually enlarging $J_2/J_1$. For $J_2/J_1\leq0.5$, the trajectory lies in the horizontal plane, $\phi=\pi$, while it jumps to $\phi=0$ plane for $J_2/J_1\geq0.5$. The $\phi=0$ plane is topologically trivial with Chern number $C=0$, whereas we find that the $\phi=\pi$ plane has both a topologically trivial ($C=0$) and a nontrivial region ($C=1$), as denoted by the yellow and grey area respectively. Interestingly, the trajectory firstly evolves into the trivial phase, then crosses the boundary into the nontrivial phase at $J_2/J_1\simeq0.3$, which persists until $J_2/J_1\simeq0.5$.

More details are included in Fig.3(b). For $0.3\gtrsim J_2/J_1$, the mean-field theory puts the system into a $\pi$-flux state with gapless Dirac cones \cite{Hsu} (see the inset to Fig.3(b)). The $\pi$-flux state describes a planar N\'{e}el AFM state after taking into account the fluctuation of the CS gauge field \cite{rui}. Remarkably, at $J_2/J_1=0.25$, when the moat begins to form (Fig.1(b)), we observe a strong growth of $\Delta\gamma$ and $A^0$. Note that the $A^0$ term breaks the sublattice symmetry while $\Delta\gamma$ violates TRS. They are two competing masses both working towards gapping out the $\pi$-flux state at small $J_2/J_1$. For $0.3\gtrsim J_2/J_1>0.25$, $A^0$ has stronger effect, giving rise to a trivial insulator state with $C=0$. Whereas, for $J_2/J_1\gtrsim 0.3$, $\Delta\gamma$ dominates and results in the band inversion that is responsible for a Chern insulator state with $C=1$. This reminds one of the celebrated Haldane model on honeycomb lattice \cite{haldanebb}; however, the state here is originated from strong frustration allowing for fractionalized quasi-particles.
Finally, we find that the Chern insulator is destabilized, towards a spin density wave (SDW) state for $J_2/J_1\ge0.50$, as a result of the emergence of a new Fermi pocket in the CS spectrum \cite{explain}.

The identified Chern insulator on square lattice for $0.5\gtrsim J_2/J_1\gtrsim0.3 $ agrees well with our tensor network calculation  in Fig.1(c): (A) It is a gapped state in bulk, ensuring the absence of in-plane magnetic ordering of any sort including $\langle S^{x,y}_{\mathbf{r}}\rangle=0$; (B) $C=1$ validates the violation of the TRS and the nonzero chirality order $\chi$; (C) The staggered CS flux $\Delta\gamma$ accurately accounts for the out-of-plane N\'{e}el Ising order;  (D) The analytical CSL region $0.5\gtrsim J_2/J_1\gtrsim0.3 $ agrees well with the numerical one $0.44\gtrsim J_2/J_1\gtrsim0.33$. These results suggest stabilization of a non-uniform CSL \cite{sedrakyanb}, supporting both semionic (flux flips) and fermionic low-energy excitations, on a square lattice.

\emph{Proximity induced d-wave superconductivity.} --
We now turn on $J_c$ and examine $H_{tot}$, with the focus on the region $0.5\gtrsim J_2/J_1\gtrsim0.3 $ where the spin system found itself in the CSL state.  The spins in $H_{sys}$ are composite objects represented by CS f-fermions and fluxes. Therefore, the spin exchange interaction, $H_c$, represents a hybridization between the environmental c-fermions and f-fermions. One can integrate out the Gaussian f-fermions in $S=S_{sys}+S_{c}$ to obtain an action describing the renormalization effect that CSL has on the fermionic bath:
\begin{equation}\label{eq8}
  \delta S=-J^2_c\int dt \sum_{\mathbf{r}\mathbf{r}^{\prime}} c^{\dagger}_{\mathbf{r}\alpha}\sigma^+_{\alpha\beta}c_{\mathbf{r}\beta}\hat{V}_{\mathbf{r},\mathbf{r}^{\prime}}c^{\dagger}_{\mathbf{r}^{\prime}\rho}\sigma^-_{\rho\sigma}c_{\mathbf{r}^{\prime}\sigma},
\end{equation}
where $\hat{V}_{\mathbf{r},\mathbf{r}^{\prime}}=\hat{G}_{\mathbf{r},\mathbf{r}^{\prime}}e^{-iA_{\mathbf{r},\mathbf{r}^{\prime}}}$, with $\hat{G}^{-1}_{\mathbf{r},\mathbf{r}^{\prime}}=(i\partial_{t}-A^0_{\mathbf{r}})\delta_{\mathbf{r},\mathbf{r}^{\prime}}-J_{\mathbf{r},\mathbf{r}^{\prime}}e^{iA_{\mathbf{r},\mathbf{r}^{\prime}}}$. The action $\delta S$ consists of a retarded on-site as well as an off-site interaction between c-fermions. The former could lead to charge orders when there is the Fermi surface nesting but this is not our focus in this work.
The latter,
\begin{equation}\label{eq11}
  V_{N}=\sum_{\mathbf{r},\mathbf{r}^{\prime}}\frac{J^2_c}{J_{\mathbf{r},\mathbf{r}^{\prime}}}e^{-2iA_{\mathbf{r},\mathbf{r}^{\prime}}}c^{\dagger}_{\mathbf{r}\uparrow}c_{\mathbf{r}^{\prime}\uparrow}c^{\dagger}_{\mathbf{r}^{\prime}\downarrow}c_{\mathbf{r}\downarrow},
\end{equation}
describes the microscopic process where a c-fermion hybridizes with a f-fermion at $\mathbf{r}$, then the f-fermion propagates to $\mathbf{r}^{\prime}$ and hybridizes with another c-fermion at $\mathbf{r}^{\prime}$. A phase factor $e^{-iA_{\mathbf{r},\mathbf{r}^{\prime}}}$ is accumulated by the f-fermion during the propagation, whereas another phase factor $e^{-iA_{\mathbf{r},\mathbf{r}^{\prime}}}$ arises during two hybridization processes, generating a total phase $e^{-2iA_{\mathbf{r},\mathbf{r}^{\prime}}}$.
\begin{figure}[tbp]
\includegraphics[width=\linewidth]{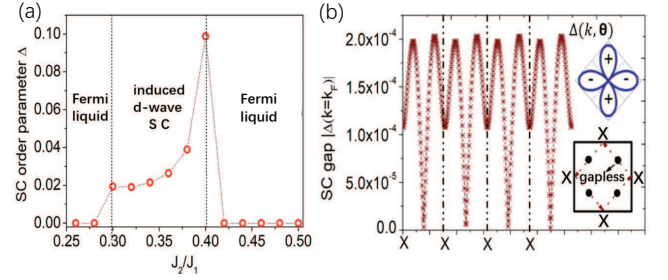}
\label{fig7}
\caption{(color online) (a) The self-consistently calculated SC order parameter $\Delta$ for $0.25<J_2/J_1<0.5$. $J_c=0.05$. (b) The calculated $|\Delta(k,\theta)|$ along the Fermi surface (the dotted square in the right lower inset). Gapless points with $|\Delta(k,\theta)|=0$ are found. The right upper inset explicitly shows the $d_{x^2-y^2}$ paring symmetry.}
\end{figure}

The interaction vertex $V_N$ can be a relevant operator responsible for the SC instability even for $J_c\rightarrow0$.
The CS phase $e^{-i2A_{\mathbf{r},\mathbf{r}^{\prime}}}$, e.g.,  $\pi/2<2A_{\mathbf{r},\mathbf{r}^{\prime}}<3\pi/2$, contributes a negative real component and therefore an attractive term between c-fermions at $\mathbf{r}$ and $\mathbf{r}^{\prime}$.
This is further quantified by making the Hubbard-Stratonovich transformation and introducing bosonic fields $\Delta_{\mathbf{r}^{\prime},\mathbf{r}}=\langle c_{\mathbf{r}^{\prime}\downarrow}c_{\mathbf{r}\uparrow}\rangle$ with respect to $V_N$.  Neglecting the spatial fluctuation of the bosonic field, we arrive at the mean-field Hamiltonian and determine the SC order parameter $\Delta$ self-consistently \cite{explain}.

The calculated $\Delta$ as a function of $J_2/J_1$ is shown in Fig.4(a). We find that a SC state with $\Delta\neq0$ is stabilized for $0.4>J_2/J_1>0.3$, where a negative condensation energy is generated \cite{explain}. We also calculate the SC gap function $|\Delta(\mathbf{k})|$ along the Fermi surface of c-fermions (inset to Fig.4(b)). Gapless nodes of $\Delta(\mathbf{k})$ with sign changes are found, which are located at the centers between X-points, as shown by Fig.4(b). The pairing symmetry can be analytically inferred from the NN pairing term \cite{explain}, $V_1\sum_{\langle \mathbf{r},\mathbf{r}^{\prime}\rangle}\Delta e^{-2iA_{ \mathbf{r},\mathbf{r}^{\prime}}}c_{\mathbf{r}\downarrow}c_{\mathbf{r}^{\prime}\uparrow}$. With the NN CS phases in Fig.2(b), one obtains a $\mathbf{k}$-dependent SC gap  $\Delta(\mathbf{k})=2\Delta(e^{i\phi}\cos k_x+\cos k_y)$. Note that $\phi=\pi$ for the whole CSL region (Fig.3(b)), the gap function is reduced to $\Delta(\mathbf{k})=2\Delta(\cos k_y-\cos k_y)$ with the standard $d_{x^2-y^2}$ symmetry.
Thus, it becomes clear that the d-wave pairing symmetry is attributed to (i) the square lattice symmetry and (ii) the $\pi$ flux state inherited from the N\'{e}el AFM order at small $J_2/J_1$.

\emph{Summary.}--We have numerically and analytically found that the TRS can be spontaneously broken in the $J_1-J_2-J_3$ AFM square lattice XY model, stabilizing a non-uniform CSL whose low-energy behavior is equivalent to a Chern insulator in terms of CS fermions on a square lattice coupled to the Chern-Simons gauge field. Based the non-uniform CSL, we reveal a proximity effect when it is coupled to nearby free-fermions, which further demonstrates a new superconducting mechanism arising from the interplay between frustrated magnets and itinerant fermions on square lattices. Furthermore, being of XY nature, the spin model under consideration is a plausible candidate to be realized using ultra-cold atoms. The work thus also opens up possibilities to study the interplay between CSLs and SCs in experiments with cold atoms  in regular optical lattices.



\begin{acknowledgments}
This work was supported by the National Key  R\&D Program of China (Grant No. 2017YFA0303200), and by NSFC under Grants No. 11574217 and No. 60825402. H.Z. acknowledges support from Science and Technology Commission of Shanghai Municipality (Grants No. 16DZ2260200).
T.A.S. acknowledges startup funds from UMass Amherst.
\end{acknowledgments}

\pagebreak
\vspace{5cm}
\widetext
\setcounter{equation}{0}
\setcounter{figure}{0}
\setcounter{table}{0}
\setcounter{page}{1}
\makeatletter
\renewcommand{\theequation}{S\arabic{equation}}
\renewcommand{\thefigure}{S\arabic{figure}}
\renewcommand{\bibnumfmt}[1]{[S#1]}
\renewcommand{\citenumfont}[1]{S#1}

\pagebreak
\vspace{5cm}
\widetext
\begin{center}
\textbf{\large Supplemental material for: D-wave superconductivity induced by proximity to the non-uniform chiral spin liquid on a square lattice}
\end{center}




\date{\today }
\maketitle
\section{Dispersion of hardcore bosons}
The spin-1/2 operators fulfill the following commutation rules, $\left[S^+_{\mathbf{r}},S^-_{\mathbf{r}^{\prime}}\right]=2\delta_{\mathbf{r},\mathbf{r}^{\prime}} S^z_{\mathbf{r}}$ and $\left[S^{\pm}_{\mathbf{r}},S^z_{\mathbf{r}^{\prime}}\right]=\mp\delta_{\mathbf{r},\mathbf{r}^{\prime}}S^{\pm}_{\mathbf{r}}$,
where $S^{\pm}_{\mathbf{r}}=S^x_{\mathbf{r}}+iS^y_{\mathbf{r}}$. Moreover, they also must satisfy the following constraint conditions: $S^{+}_{\mathbf{r}}S^-_{\mathbf{r}}+S^-_{\mathbf{r}}S^+_{\mathbf{r}}=1$ and $(S^+_{\mathbf{r}})^2=(S^-_{\mathbf{r}})^2=0$. One can introduce a set of bosonic operators that exactly satisfy all the above conditions, i.e., $S^+_{\mathbf{r}}=b_{\mathbf{r}}$, $S^-_{\mathbf{r}}=b^{\dagger}_{\mathbf{r}}$, and $S^z_{\mathbf{r}}=1/2-b^{\dagger}_{\mathbf{r}}b_{\mathbf{r}}$. In order to satisfy the on-site exclusion rule of spin operators, one still needs to require the hardcore condition for the bosons with the occupation number per site to be restricted to either 0 or 1. This is achieved by introducing an infinite on-site Hubbard repulsion with $U\rightarrow\infty$.

Using the hardcore representation, we can map the spin-1/2 $J_1-J_2-J_3$ AFM XY model to a bosonic Hamiltonian. The single-particle sector without considering the Hubbard repulsion then reads as
\begin{equation}\label{eqsup1}
  H_{boson}=\frac{1}{2}\sum_{\mathbf{r},\mathbf{r}^{\prime}}J_{\mathbf{r},\mathbf{r}^{\prime}}(b^{\dagger}_{\mathbf{r}}b_{\mathbf{r}^{\prime}}+h.c.),
\end{equation}
where the interaction $J_{\mathbf{r},\mathbf{r}^{\prime}}$ is defined up to the NNNN interaction, the same as in the main text. The single-particle dispersion of the hardcore bosons can be readily obtained by diagonalization of Eq.\eqref{eqsup1}.

\section{Scaling of the chirality order}
We calculate the scaling of the chirality order with respect to $\beta$ and plot it in the following figure. For the CSL region $0.44\gtrsim J_2/J_1\gtrsim 0.33$ (Figure 1a), the scaling behavior is found to be nearly linear, which is an inherent feature of CSL. The magnitude of $\chi$ in the CSL region is much larger than that in the magnetic order region, which is vanishingly small ($\sim 10^{-5}$)in the latter, see Figure 1b.
\begin{figure}[bp]
\includegraphics[width=\linewidth]{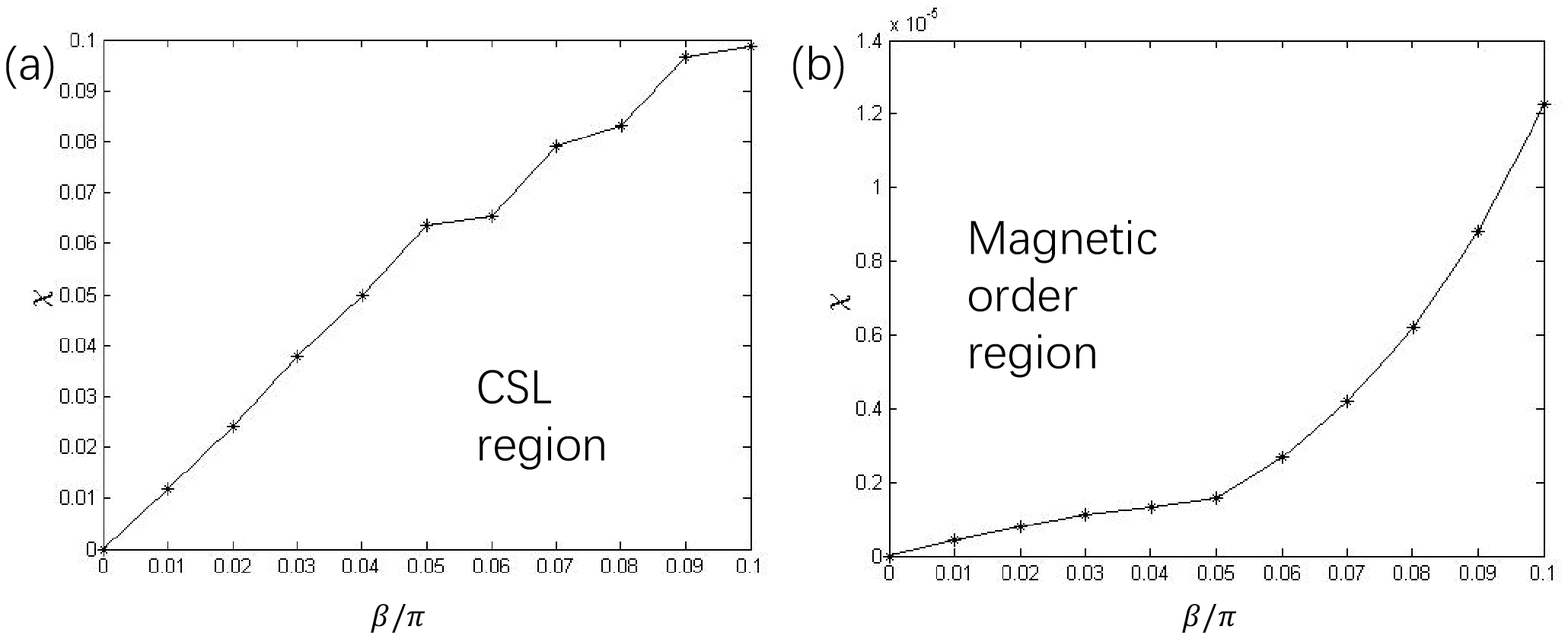}
\label{fig5}
\caption{(color online) Scaling of the chirality order as a function of $\beta$ in (a) the CSL region (b) the magnetic order region.}
\end{figure}

\section{Mean-field theory of the non-uniform CSL}
Inserting the CS representation to $H_{sys}$, the Hamiltonian is cast in the form,
\begin{equation}\label{eq4sup}
\begin{split}
  H_{sys}&=J_1\sum_{\mathbf{r}j}f^{\dagger}_{\mathbf{r}}f_{\mathbf{r}+\mathbf{e}_j}e^{i\sum_{\tilde{\mathbf{r}}}\phi_{\tilde{\mathbf{r}};\mathbf{r},\mathbf{r}+\mathbf{e}_j}n_{\tilde{\mathbf{r}}}}
  +J_2\sum_{\mathbf{r}j}f^{\dagger}_{\mathbf{r}}f_{\mathbf{r}+\mathbf{a}_j}e^{i\sum_{\tilde{\mathbf{r}}}\phi_{\tilde{\mathbf{r}};\mathbf{r},\mathbf{r}+\mathbf{a}_j}n_{\tilde{\mathbf{r}}}}
  +J_3\sum_{\mathbf{r}j}f^{\dagger}_{\mathbf{r}}f_{\mathbf{r}+\mathbf{b}_j}+h.c.,
\end{split}
\end{equation}
where we define $\mathbf{e}_j$, $\mathbf{a}_j$ and $\mathbf{b}_j$ as NN, NNN and NNNN bond vectors. $\phi_{\tilde{\mathbf{r}};\mathbf{\mathbf{r}},\mathbf{\mathbf{r}}+\mathbf{e}_j}$ is defined as $\phi_{\tilde{\mathbf{r}};\mathbf{\mathbf{r}},\mathbf{\mathbf{r}}+\mathbf{e}_j}=\mathrm{arg}(\mathbf{r}-\tilde{\mathbf{r}})-\mathrm{arg}(\mathbf{r}+\mathbf{e}_j-\tilde{\mathbf{r}})$ , i.e., the angle of the bond $\langle \mathbf{r},\mathbf{r}+\mathbf{e}_j\rangle$ seen by the site at $\tilde{\mathbf{r}}$, similarly for $\phi_{\tilde{\mathbf{r}};\mathbf{r},\mathbf{r}+\mathbf{a}_j}$ and $\phi_{\tilde{\mathbf{r}};\mathbf{r},\mathbf{r}+\mathbf{b}_j}$.  The CS phase in a loop centered at site $\mathbf{r}$ can be obtained as  $B_{\mathbf{r}}=\phi_{loop}=2\pi n_{\mathbf{r}}$. A Lagrangian multiplier $A^0_{\mathbf{r}}$ is then introduced to ensure this condition, leading to the path integral action describing the spinless CS fermions coupled to fluctuating CS gauge field,
\begin{equation}\label{eq5sup}
\begin{split}
  S_{sys}&=\int dt[\sum_{\mathbf{r}}f^{\dagger}_{\mathbf{r}}(i\partial_t-A^0_{\mathbf{r}})f_{\mathbf{r}}+\frac{1}{2\pi}\sum_{\mathbf{r}}A^0_{\mathbf{r}}B_{\mathbf{r}}
  -J_1\sum_{\mathbf{r}j}f^{\dagger}_{\mathbf{r}}f_{\mathbf{r}+\mathbf{e}_j}e^{iA_{\mathbf{r},\mathbf{r}+\mathbf{e}_j}}-J_2\sum_{\mathbf{r}j}f^{\dagger}_{\mathbf{r}}f_{\mathbf{r}+\mathbf{a}_j}e^{iA_{\mathbf{r},\mathbf{r}+\mathbf{a}_j}}\\
  &-J_3\sum_{\mathbf{r}j}f^{\dagger}_{\mathbf{r}}f_{\mathbf{r}+\mathbf{b}_j}-h.c.],
\end{split}
\end{equation}
where we have defined the gauge field on bond as $A_{\mathbf{r},\mathbf{r}+\mathbf{e}_j}=\sum_{\tilde{\mathbf{r}}}\phi_{\tilde{\mathbf{r}},\mathbf{r}\mathbf{r}+\mathbf{e}_j}n_{\tilde{\mathbf{r}}}$. $A^0_{\mathbf{r}}$ is the Lagrangian multiplier field introduced to satisfy the ``Stocks theorem" of CS gauge fields.

The above action is bilinear in terms of CS fermions $f_{\mathbf{r}}$. After integrating out the CS fermions exactly, the fermionic free energy functional can be obtained as $S_{eff}=W[A^0_{\mathbf{r}},\phi,\gamma]+\frac{1}{2\pi}\sum_{\mathbf{r}}A^0_{\mathbf{r}}B_{\mathbf{r}}[\phi,\gamma]$, where $B_{\mathbf{r}}$ is the field that depends on the CS flux $\phi$ and $\gamma$. The mean-field equations are obtained by taking variation over $A^0_{\mathbf{r}}$ and $B_{\mathbf{r}}$, i.e., $\delta_{A^0_{\mathbf{r}}}S_{eff}=0$ and $\delta_{B_{\mathbf{r}}}S_{eff}=0$.
\begin{figure}[tbp]
\includegraphics[width=\linewidth]{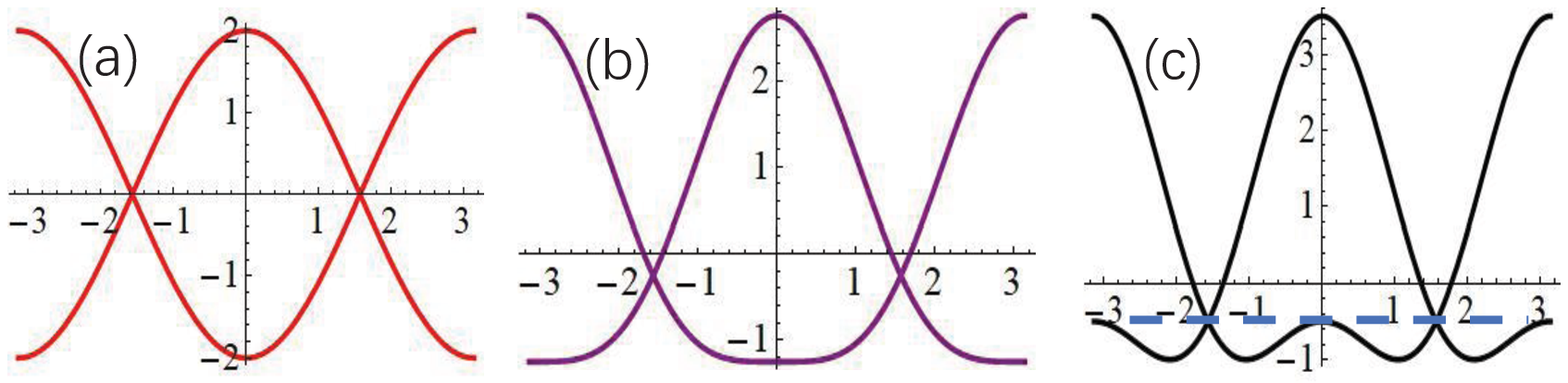}
\label{fig5}
\caption{(color online) The CS fermion dispersion along $k_y=k_x$ with turning off all the CS gauge fields at (a) $J_2/J_1=0$ (b) $J_2/J_1=0.25$ and (c) $J_2/J_1=0.5$.}
\end{figure}
The flux configuration of Fig.~2 of the main text should be taken into account in the above action. With gauging out the symmetric component of the gauge field, which is gauge equivalent to zero due to the periodicity of $2\pi$, the term  $A^0_{\mathbf{r}}$ becomes, i.e., $\frac{1}{2\pi}\sum_{\mathbf{r}}A^0_{\mathbf{r}}B_{\mathbf{r}}=\frac{N}{2\pi}A^0\Delta\gamma$, since $B^A_{\mathbf{r}}-B^B_{\mathbf{r}}=\gamma_2-\gamma_1=\Delta\gamma$ as shown by Fig.2(a). Moreover, the term $\sum_{\mathbf{r}}f^{\dagger}A^0_{\mathbf{r}}f_{\mathbf{r}}$ becomes a staggered chemical potential that breaks sublattice symmetry, i.e., $\sum_{\mathbf{r}}f^{\dagger}A^0_{\mathbf{r}}f_{\mathbf{r}}=\sum_{\mathbf{r}}[f^{\dagger}_{\mathbf{r}A}A^0f_{\mathbf{r}A}-f^{\dagger}_{\mathbf{r}B}A^0f_{\mathbf{r}B}]$.

Here, we also plot the dispersion of the CS fermions (along $k_y=k_x$) with turning off all the gauge fields. As shown in Figure 2 above,  with increasing  $J_2/J_1$ from $0$ (Fig.2(a)) to $J_2/J_1=0.25$, the dispersion at the energy bottom becomes more and more flat. After $J_2/J_1>0.25$ a moat-shape band is formed around $\Gamma$ point, corresponding to the moat band in the hardcore boson representation. Interestingly, when $J_2/J_1=0.5$, a new Fermi pocket begins to emerge at the Fermi surface (around the $\Gamma$ point) as shown by Fig.2(c). This is the reason for the instability to a spin density wave (SDW) order for $J_2/J_1>0.5$.

\section{Superconducting instability of the fermion bath}
After turning on $J_c$, we have shown that the non-uniform CSL generates effective interactions between fermions. The total effective Hamiltonian describing the fermionic environment then becomes,

\begin{equation}\label{eq12sup}
\begin{split}
  H&=-t\sum_{\langle \mathbf{r},\mathbf{r}^{\prime}\rangle}c^{\dagger}_{\mathbf{r}\sigma}c_{\mathbf{r}^{\prime}\sigma}+H.C.
  +V_{1}\sum_{\langle\mathbf{r},\mathbf{r}^{\prime}\rangle}e^{-2iA_{\mathbf{r},\mathbf{r}^{\prime}}}c^{\dagger}_{\mathbf{r}\uparrow}c^{\dagger}_{\mathbf{r}^{\prime}\downarrow}c_{\mathbf{r}\downarrow}c_{\mathbf{r}^{\prime}\uparrow}
  +V_{2}\sum_{ \langle\langle\mathbf{r},\mathbf{r}^{\prime}\rangle\rangle}e^{-2iA_{\mathbf{r},\mathbf{r}^{\prime}}}c^{\dagger}_{\mathbf{r}\uparrow}c^{\dagger}_{\mathbf{r}^{\prime}\downarrow}c_{\mathbf{r}\downarrow}c_{\mathbf{r}^{\prime}\uparrow}\\
  &+V_{3}\sum_{\langle\langle\langle\mathbf{r},\mathbf{r}^{\prime}\rangle\rangle\rangle}c^{\dagger}_{\mathbf{r}\uparrow}c^{\dagger}_{\mathbf{r}^{\prime}\downarrow}c_{\mathbf{r}\downarrow}c_{\mathbf{r}^{\prime}\uparrow},
\end{split}
\end{equation}
where $V_i=J^2_c/J_i$ with $i=1,2,3$. $A_{\mathbf{r},\mathbf{r}^{\prime}}$ are the CS gauge phases inherited from the CSL state, as shown by Fig.2 in the main text. Then we solve the Hamiltonian using the self-consistent mean-field theory.  In the particle-particle channel, we introduce Hubbard-Stratonovich bosonic field, $\Delta_{\mathbf{r}^{\prime},\mathbf{r}}=\langle c_{\mathbf{r}^{\prime}\downarrow}c_{\mathbf{r}\uparrow}\rangle$ for pairing term
on the neighbour bonds $\langle \mathbf{r},\mathbf{r}^{\prime}\rangle$.

Before proceeding, we note that the $V_3$ term is different in nature from $V_1$ and $V_2$ due to the absence of the NNNN CS gauge field of the CSL state. Hence, the $V_3$ coupling between $c$-fermions is simply a repulsive exchange interaction. Without the CS gauge field, the interaction is always repulsive between NNNN c-fermions and unable to generate any SC instability. Indeed, if one introduces the Hubbard-Stratonovich field $\Delta_{3\mathbf{r}^{\prime},\mathbf{r}}=\langle c_{\mathbf{r}^{\prime}\downarrow}c_{\mathbf{r}\uparrow}\rangle$ for NNNN pairing with $\mathbf{r}$, $\mathbf{r}^{\prime}\in \langle\langle\langle \mathbf{r}$, $\mathbf{r}^{\prime}\rangle\rangle\rangle$, one will not be able to find any nontrivial solutions with $\Delta_{3\mathbf{r}^{\prime},\mathbf{r}}\neq0$. As we focus on the SC instability of Eq.\eqref{eq12sup}, the effect of $V_3$ term becomes negligible.
Further neglecting the fluctuation of the Hubbard-Stratonovich field, we obtain the following mean-field Hamiltonian,

\begin{equation}\label{eq13sup}
\begin{split}
  H&=-t\sum_{\langle \mathbf{r},\mathbf{r}^{\prime}\rangle}c^{\dagger}_{\mathbf{r}\sigma}c_{\mathbf{r}^{\prime}\sigma}+H.C.+V_1\sum_{\langle \mathbf{r},\mathbf{r}^{\prime}\rangle}\Delta e^{-2iA_{ \mathbf{r},\mathbf{r}^{\prime}}}c_{\mathbf{r}\downarrow}c_{\mathbf{r}^{\prime}\uparrow}
  +H.C.+V_2\sum_{\langle\langle \mathbf{r},\mathbf{r}^{\prime}\rangle\rangle}\Delta e^{-2iA_{ \mathbf{r},\mathbf{r}^{\prime}}}c_{\mathbf{r}\downarrow}c_{\mathbf{r}^{\prime}\uparrow}+H.C.\\
  &-V_1\Delta^2\sum_{\langle \mathbf{r},\mathbf{r}^{\prime}\rangle}e^{2iA_{ \mathbf{r},\mathbf{r}^{\prime}}}-V_2\Delta^2\sum_{\langle\langle \mathbf{r},\mathbf{r}^{\prime}\rangle\rangle}e^{-2iA_{ \mathbf{r},\mathbf{r}^{\prime}}}.
\end{split}
\end{equation}

Upon inserting the CS flux configuration of Fig.2 of the main text,  transforming the Hamiltonian to the lattice momentum space, and then minimizing the free energy, one
will self-consistently determine the order parameters $\Delta$ for different $V_i$.
\begin{figure}[tbp]
\includegraphics[width=0.7\linewidth]{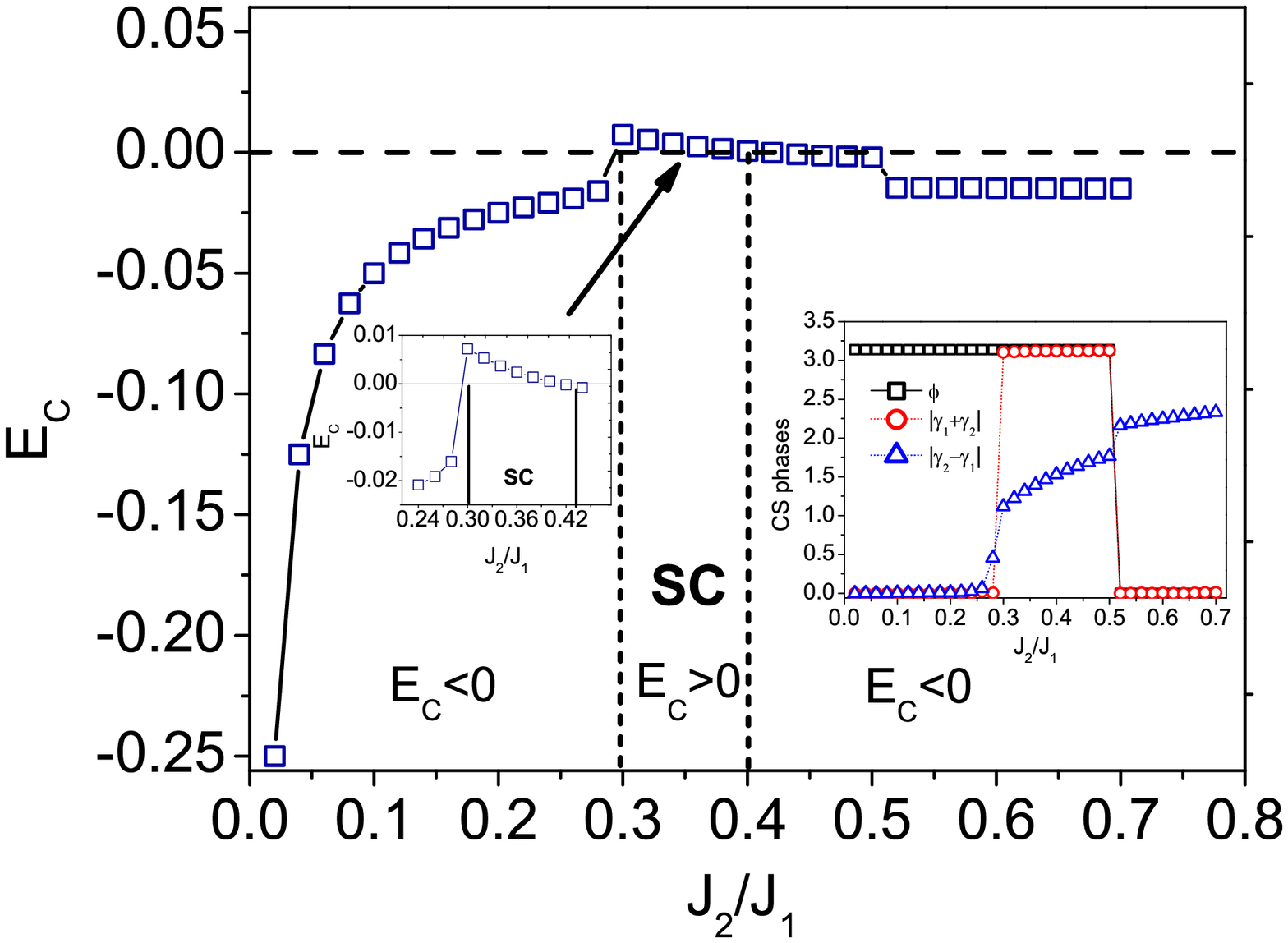}
\label{fig5}
\caption{(color online) Plot of calculated $E_c$ as a function of $J_2/J_1$. A region $0.3\leq J_2/J_1\leq0.4$ is found where $E_c>0$, stabilizing a SC order in the fermionic environment. The left inset shows the details for the $E_c>0$ region, and the right inset shows the self-consistent values of $\phi$, $|\gamma_2+\gamma_1|$ and $|\gamma_2-\gamma_1|$.}
\end{figure}

Interestingly, after taking into account the fluxes, the two constant terms of Eq.\eqref{eq13sup} become $-V_1\Delta^2\sum_{\langle \mathbf{r},\mathbf{r}^{\prime}\rangle}e^{2iA_{ \mathbf{r},\mathbf{r}^{\prime}}}=-4NV_1\Delta^2(1+\cos\phi)$ and $-V_2\Delta^2\sum_{\langle\langle \mathbf{r},\mathbf{r}^{\prime}\rangle\rangle}e^{-2iA_{ \mathbf{r},\mathbf{r}^{\prime}}}=-2NV_2\Delta^2\cos(\gamma_1+\gamma_2)\cos(\gamma_2-\gamma_1)$. Let us compare them with that of the standard BCS theory. Based on the reduced BCS Hamiltonian with a negative interaction $-V_{eff}\sum_{\mathbf{k}\mathbf{k}^{\prime}}c^{\dagger}_{\mathbf{k}^{\prime}}c^{\dagger}_{-\mathbf{k}^{\prime}}c_{-\mathbf{k}}c_{\mathbf{k}}$, and $V_{eff}>0$, a constant term is obtained as $N\Delta^2/V_{eff}$. When the coefficient $N/V_{eff}$ is larger than zero, the standard mean-field theory always gives a self-consistent nonzero solution of $\Delta$ as there forms a free energy minimum at $\Delta\neq0$ (if $N/V_{eff}<0$, it means the interaction term $-V_{eff}\sum_{\mathbf{k}\mathbf{k}^{\prime}}c^{\dagger}_{\mathbf{k}^{\prime}}c^{\dagger}_{-\mathbf{k}^{\prime}}c_{-\mathbf{k}}c_{\mathbf{k}}$ is repulsive and therefore no SC order will be generated). Comparing with the theory here, we find the following correspondence,
\begin{equation}\label{eq14sup}
  \frac{1}{V_{eff}}\sim  E_c=-4V_1(1+\cos\phi)-2V_2\cos(\gamma_1+\gamma_2)\cos(\gamma_2-\gamma_1).
\end{equation}
Through the above correspondence, one can conclude that for $E_c>0$ there will be a nontrivial SC order developed from the mean-field Hamiltonian Eq.\eqref{eq13sup}, while $\Delta=0$ for $E_c\le0$. Furthermore, $E_c$ is a function of the CS gauge field $\phi$, $\gamma_1$ and $\gamma_2$, which have been self-consistently determined (see main text). As shown by Fig.3 above, we find a region $0.3\leq J_2/J_1\leq0.4$ with $E_c>0$ (the left inset), stabilizing a nontrivial SC order in the fermionic environment. Furthermore, for the CSL region $0.3<J_2/J_1<0.5$, we obtain $\phi=\pi$ and $|\gamma_1+\gamma_2|=\pi$ as shown by the right inset to Fig.3 above, Eq.\eqref{eq14sup} is thus reduced to $\frac{1}{V_{eff}}\sim  E_c=2V_2\cos(\gamma_2-\gamma_1)$. Therefore, we analytically find out the transition point between the Fermi liquid and the superconductivity, i.e., $|\gamma_2-\gamma_1|=\pi/2$, in the mean-field level.



\begin{thebibliography}{99}
\bibitem{footnote-here} Although a perfectly flat moat stabilizes at $J_3=J_2/2$, an approximately flat moat-like band structure emerges in a large parameter range near $J_3\sim J_2/2$. Without losing generality, we focus on $J_3=J_2/2$ in this work.
\bibitem{footnote}  A detailed numerical and analytical study on the groundstate of the spin model will be presented by a susequent work.
\bibitem{explain2} A word of caution here, however, is in order. Although the anomalous response to a week TRS breaking perturbation is a clear indication of stabilization of chirality, our numerical computations do not capture finite $\chi$ exactly at $\beta=0$ due to the limited D-bond, which is not able to fully capture an inital state with long-range quantum entanglement. DMRG calculations on finite cylinder can demonstrate the subtle point clearly.
\end{thebibliography}
\end{document}